# DNA-assembled nanoarchitectures with multiple components in regulated and coordinated motion

Pengfei Zhan[1#], Maximilian J. Urban[1,2#], Steffen Both[3], Xiaoyang Duan[1,2], Anton Kuzyk[4], Thomas Weiss[3], and Na Liu[1,2,*]

[1]*Max Planck Institute for Intelligent Systems, Heisenbergstrasse 3, D-70569 Stuttgart, Germany*
[2]*Kirchhoff Institute for Physics, Heidelberg University, Im Neuenheimer Feld 227, D-69120 Heidelberg, Germany*
[3]*4th Physics Institute and Stuttgart Research Center of Photonic Engineering, University of Stuttgart, 70569 Stuttgart, Germany*
[4]*Department of Neuroscience and Biomedical Engineering, Aalto University, School of Science, P.O. Box 12200, FI-00076 Aalto, Finland*

*e-mail: na.liu@kip.uni-heidelberg.de*

**Coordinating functional parts to operate in concert is essential for machinery. In gear trains, meshed gears are compactly interlocked, working together to impose rotation or translation. In photosynthetic systems, a variety of biological entities in the thylakoid membrane interact with each other, converting light energy into chemical energy. However, coordinating individual parts to carry out regulated and coordinated motion within an artificial nanoarchitecture poses great challenges, owing to the requisite control on the nanoscale. Here, we demonstrate DNA-directed nanosystems, which comprise hierarchically-assembled DNA origami filaments, fluorophores, and gold nanocrystals. These individual building blocks can execute independent, synchronous, or joint motion upon external inputs. The dynamic processes are *in situ* optically monitored using fluorescence spectroscopy, taking advantage of the sensitive distance-dependent interactions between the gold nanocrystals and fluorophores positioned on the DNA origami. Our work leverages the complexity of DNA-based artificial nanosystems with tailored dynamic functionality, representing a viable route towards technomimetic nanomachinery.**

Machines are built with multiple parts, which can work together for a particular function. In biological machinery as complex as photosynthetic systems[1], membrane proteins, pigments, chlorophyll, and others interact and work with each other, efficiently capturing light and converting it to chemical energy[2–4]. In manmade machinery such as gear trains (see Fig. 1a), macroscopic elements including gears, shafts, and racks are assembled together to transmit power and alter rotary speeds. For instance, in an epicyclic gearset, two planet gears (grey) mounted on a single shaft or different shafts can revolve around a centrally-pivoted sun gear (brown) synchronously or independently. In a rack-and-pinion gearset, a pinion gear (brown) is positioned in between two steering racks (grey). As the pinion gear turns, it slides the racks either to the left or right, transforming rotation into linear motion.

Such seemingly straightforward coordination and control over motion among multiple components become quite intriguing, if one attempts to build their analogues in the nanoscopic world. The main reason for this is twofold. First, the spatial arrangements of distinct components within a nanoarchitecture is crucial for its functionality. However, hierarchical assembly methods that allow for organizing nanoscale objects with precise compositions, positions, and relative orientations in a modular fashion are few[5]. Second, to mimic the dynamic complexity of macroscopic systems, proper external inputs with accurate spatiotemporal control have to be imposed, so that different components within the nanoarchitectures can act in response, carrying out regulated and coordinated motion. Nevertheless, this requires remarkable programmability and addressability of the nanoarchitectures, which have to be endowed during the assembly process.

The DNA origami technique is ideally suited to provide the solution. The formation of DNA origami involves the folding of a long DNA scaffold by hundreds of short staple strands into nanostructures with nearly arbitrary shapes[6,7]. As each staple strand possesses a unique sequence with a deterministic position in the formed structure, DNA origami is a fully addressable 'molecular pegboard' to organize discrete objects in three dimensions with nanoscale accuracy[8–14]. Most importantly, this technique imparts dynamic programmability to the assembled nanoarchitectures, in which the individual components can carry out regulated motion and exhibit programmable immediate conformational changes upon a variety of external inputs[15–24]. Although there is still a long way ahead to directly transfer the functional characteristics of macroscopic devices and build their technomimetic analogues, any attempts along this direction will provoke valuable insights into how functional structures have to be built and how the communication among multiple components can be mastered on the nanoscale[25].

Figure 1b illustrates the concept of DNA-assembled multicomponent dynamic nanosystems. Nanoscopic elements such as DNA origami filaments, fluorophores, and metallic nanocrystals including gold nanorods (AuNRs) or spherical nanoparticles (AuNPs) are assembled together to form dynamic nanoarchitectures, in which the individual components can execute regulated and coordinated motion with high fidelity.

In the following, we will sequentially elucidate three examples, which yield independent, synchronous, or joint motion.

Figure 2a shows an epicyclic gearset, in which two planet gears (grey) of different diameters are mounted on a sun gear (brown) using two shafts. The two planet gears can independently revolve around the sun gear by rotation. Figure 2b illustrates a self-assembled hybrid system, in which two DNA origami filaments of different diameters are anchored on the side-surface of a AuNR. They can independently revolve around the AuNR by rotation powered by DNA fuels. The small (A, 13-helix) and large (B, 23-helix) origami filaments formed by folding M13 scaffold, staples, and foothold strands through a hierarchical assembly process are linked to a AuNR (35 nm in length and 10 nm in diameter) by DNA hybridization. Filaments A and B are connected at their ends using the scaffold strand to ensure a correct relative orientation and meanwhile allow for structural flexibility. As shown by the cross-section view of the system in Fig. 2c, three rows of footholds (coded 1–3) evenly separated by 120° are extended from filament A, whereas six rows of footholds (coded 4–9) evenly separated by 60° are extended from filament B. There are five binding sites with identical footholds in each row. The design details can be found in Supplementary Fig. 1 and Tables 1 and 2. Filaments A and B are anchored on the AuNR with four foothold rows, two from each. As shown in Fig. 2c, they are first bound to the AuNR with foothold rows 1, 2 and 4, 5, respectively. The rest of the foothold rows are deactivated using respective blocking strands. The revolution of filament A around the AuNR starts upon simultaneous addition of blocking strands 2′ and removal strands $\bar{3}$, while filament B anchors on the AuNR. The blocking strands detach foothold row 2 from the AuNR and subsequently block it through toehold-mediated strand displacement reactions[26]. Meanwhile, the removal strands activate foothold row 3 through toehold-mediated strand displacement reactions so that it can subsequently bind to the AuNR. As a result, filament A carries out a 120° rotation counterclockwise around the side-surface of the AuNR. It is noteworthy that the rotation direction is fully programmable, depending on the DNA fuels added in the system. For instance, a clockwise rotation of filament A can be imposed upon addition of blocking strands 1′ and removal strands $\bar{3}$. Similarly, filament B can independently revolve around the side-surface of the AuNR by adding corresponding DNA fuels, resulting in a 60° rotation clockwise or counterclockwise each step.

Figure 2d presents the transmission electron microscopy (TEM) image of the assembled AuNR-origami nanostructures before rotation. The overview TEM image confirms that in the individual structures two DNA origami filaments of different diameters are attached to one AuNR. The averaged TEM image (see the inset image in Fig. 2d) demonstrates an excellent structural homogeneity. More TEM images can be found in Supplementary Figs 2–4. To optically monitor the independent revolution process, two fluorophores (ATTO 550 and ATTO 647N) are tethered on filaments A and B along their long axes, respectively (see Supplementary Fig. 5)[27]. When the two filaments revolve independently around the side-surface of the AuNR, the distance-dependent interactions of ATTO 550 and ATTO 647N with the AuNR give rise to *in* *situ* fluorescence intensity changes, readily correlating spatial information with optical responses.

Figure 2e illustrates a representative route for an independent revolution process with nine distinct states (I–IX). The triangle and hexagon shapes symbolize filaments A and B, respectively, in resemblance to their foothold row arrangements. The grey disk represents the cross-section view of the AuNR. In order to identify the rotation directions and meanwhile simplify the illustrations, two foothold rows 1, 2 and 4, 5 from filaments A and B, respectively, are labeled in each plot. The positions of the two fluorophores and their relative distances to the AuNR surface along the respective radial directions are also given for each state. The revolution process was *in situ* monitored using the dual-wavelength time-scan function of a fluorescence spectrometer (Jasco-FP8500) at two emission wavelengths of 578 nm and 663 nm with excitation wavelengths of 550 nm and 647 nm, respectively. The measured fluorescence data are presented in Fig. 2f. Following I–III, filament B rotates 60° each step counterclockwise around the AuNR upon addition of the corresponding DNA fuels, while filament A keeps its anchoring position on the AuNR (see Supplementary Table 3). Such successive dynamic behavior is directly correlated with fluorescence intensity changes of ATTO 647N (red line) in real time as shown in Fig. 2f. More specifically, the distance enlargement (from I to II) and reduction (from II to III) between ATTO 647N and the AuNR lead to fluorescence intensity increase and decrease, respectively (see the detailed structural parameters in Supplementary Fig. 6a, b). As the distance between ATTO 550 and the AuNR remains unchanged, its fluorescence intensity stays constant during I–III (see the blue line in Fig. 2f). Subsequently, filaments A and B carry out simultaneous, yet independent revolution counterclockwise around the AuNR following IV–VI, driven by their respective DNA fuels. This results in distinct fluorescence intensity changes for both ATTO 550 and ATTO 647N as shown in Fig. 2f. Interestingly, a kink (see the black-dashed frame) in fluorescence is observed, when filament A transits from IV to V. It arises from the fact that during this single step ATTO 550 first experiences a distance increase and then a decrease with respect to the AuNR (see the details in Supplementary Fig. 7). Such spatial information of subtle distance changes on the nanoscale is optically visualized in real time, thanks to the high sensitivity of fluorescence spectroscopy. Following VII–IX, filament B remains at its anchoring position, while filament A revolves around the AuNR counterclockwise upon addition of the corresponding DNA fuels. This gives rise to fluorescence intensity changes of ATTO 550 and intensity invariance of ATTO 647N, respectively. Another kink of the same origin as discussed above is observed for ATTO 550, when filament A transits from VII to VIII. Additional experimental data, demonstrating the high fidelity and reversibility of the system are shown in Supplementary Fig. 8.

To quantitatively corroborate the experimental observations, theoretical calculations have been performed as shown in Fig. 2g. The fluorescence intensity changes of the fluorophores owing to their distance-dependent interactions with the AuNR are calculated from[28]

$$\frac{\gamma_{\text{fl}}}{\gamma_{\text{fl},0}} = \frac{q}{q_0}\frac{\gamma_{\text{exc}}}{\gamma_{\text{exc},0}}, \quad (1)$$

in which $\gamma_{fl}$, $q$, and $\gamma_{exc}$ denote the fluorescence rate, quantum yield, and excitation rate in the presence of the AuNR. $\gamma_{fl,0}$, $q_0$, and $\gamma_{exc,0}$ describe the same quantities, but for the free fluorophores. The excitation enhancement $\gamma_{exc}/\gamma_{exc,0}$ can be derived from the electromagnetic near-fields generated by the AuNR, when interacting with a plane wave at the excitation wavelength of 550 nm for ATTO 550 or 647 nm for ATTO 647N. To account for the random orientations of the individual structures in solution, the excitation enhancement for each fluorophore is averaged over all possible directions and polarizations.

In equation (1), the quantum yield $q$ is given as[28]

$$q = \frac{\gamma_r/\gamma_{r,0}}{\gamma_r/\gamma_{r,0} + \gamma_{abs}/\gamma_{r,0} + (1-q_0)/q_0}, \qquad (2)$$

where $\gamma_r$ denotes the fluorescence rate in the presence of the AuNR, $\gamma_{abs}$ is the rate of energy absorption in the AuNR, and $\gamma_{r,0}$ is the intrinsic radiative decay rate of the free fluorophore. The factors $\gamma_r/\gamma_{r,0}$ and $\gamma_{abs}/\gamma_{r,0}$ are deduced from the electromagnetic near-field simulations of an emitting dipole placed next to the AuNR. The emission spectra of the fluorophores are taken into account by averaging $\gamma_r/\gamma_{r,0}$ and $\gamma_{abs}/\gamma_{r,0}$ over the intrinsic emission spectra of the free fluorophores[29]. To account for the rotational freedom of the fluorophores, $\gamma_r/\gamma_{r,0}$ and $\gamma_{abs}/\gamma_{r,0}$ are averaged over all possible dipole orientations. Calculation details can be found in Supplementary Notes and Supplementary Figs 9–11. As shown by Fig. 2f, g, the experimental and calculated results agree very well.

Figure 3a presents another epicyclic gearset, in which two planet gears (grey) of the same size are mounted on a sun gear (brown) using one single shaft. The two planet gears can synchronously revolve around the sun gear by rotation. Correspondingly, a self-assembled hybrid system is shown in Fig. 3b, in which two 23-helix DNA origami filaments of the same size are anchored on the side-surface of a AuNR. The detailed structural design and parameters can be found in Supplementary Figs 12 and 13. As shown in Fig. 3c, six rows of footholds (coded 1–6) evenly separated by 60° are extended from each filament. The two filaments are bound to the AuNR with the same two rows of footholds, so that they can be driven synchronously to revolve around the AuNR clockwise or counterclockwise using the same set of blocking and removal strands. Figure 3d presents the TEM image of the assembled structures. The thinner 6-helix tails at the end of the filaments (see Supplementary Fig. 12) are designed to identify their relative orientation and more importantly to resolve the rotation steps to some extent. For instance, the averaged TEM image in Fig. 3d corresponds to state I in Fig. 3e, in which the two tails are both aligned close to the AuNR (see also Supplementary Figs 14 and 15). When the two filaments synchronously revolve around the AuNR following I–VI, both filaments rotate counterclockwise by 60° in each step, as shown in Fig. 3c, powered by the same set of DNA fuels. Due to synchronous motion, filaments A and B together with the AuNR remain coaxially in the cross-section plane (see Fig. 3e). The averaged TEM image at state IV is shown in the inset of Supplementary Fig. 16, in which the two tails are both aligned away from the AuNR (see also Supplementary Fig. 17). The two filaments can also carry out clockwise rotation synchronously upon addition of the corresponding fuels (see Supplementary Table 4). Figure 3f shows the *in situ* fluorescence intensity tracking of ATTO 550 and ATTO 647N, respectively, during the synchronous revolution process. The six distinct states can be optically resolved well. The experimental observations are in good agreement with theoretical calculations as shown in Fig. 3g. The reversibility of the system is demonstrated in Supplementary Fig. 18.

Taking a step further, synchronous revolution from epicyclic gearing and relative sliding from rack-and-pinion gearing are combined within one complex DNA-assembled nanosystem, yielding joint motion. A rack-and-pinion gearset, comprising double racks and one gear (see Fig. 4a) converts rotational motion into linear motion. The circular gear, called a pinion spins in between two linear gears, called racks. The rotational motion imposed by the pinion causes the racks to move relative to each other. Figure 4b shows a self-assembled system, in which two 23-helix DNA origami filaments of the same size are anchored on a AuNP of 10 nm in diameter. In total, twelve rows of footholds (see Supplementary Fig. 19) are extended from the filaments to enable synchronous revolution and relative sliding along two different double-racks. The foothold row arrangement is depicted by the side-view and cross-section-view of the system in Fig. 4c. The two filaments are bound to the AuNP with four foothold rows, two from each. For instance, at state I in Fig. 4d, filaments A and B are anchored to the AuNP with foothold rows 1 and 2 (see the side-view in Fig. 4c) through DNA hybridization. The foothold rows are spaced by 7 nm along the long axis of each filament. At state I, the initial displacement between the two filaments along the long-axis direction is -14 nm. Upon addition of blocking strands 1′ and removal strands $\bar{3}$ and $\bar{4}$, the AuNP spins in between the two filaments and subsequently binds to foothold rows 3, and 4 through toehold-mediated strand displacement reactions (see Supplementary Table 5). This reduces the displacement between the two filaments to 0 nm (see II in Fig. 4d). Figure 4e presents the TEM image of the assembled nanostructures at state I (see also Supplementary Fig. 20). Due to a larger structural flexibility compared to those assembled with AuNRs, the nanostructures assembled with AuNPs show higher structural deformations after drying on the TEM grid.

Figure 4d illustrates the seven distinct states in a joint motion process. The distances of ATTO 550 and ATTO 647N relative to the AuNP surface along the respective radial directions are indicated accordingly for each state. This complex nanosystem can carry out relative sliding along two independent double-racks (see I in Fig. 4d). One is defined by foothold rows 5,4,2,1 and 1,2,3,6 located along two parallel racks in red, respectively. The other is defined by foothold rows 12,9,10 and 11,8,12 located along two parallel racks in grey, respectively. By sliding one step further, the system reaches state III from II, introducing a 14 nm displacement between the two filaments. Subsequently, the system starts synchronous revolution, departing from state III to VI in three steps (see Fig. 4d). At each step, both filaments rotate by 60° counterclockwise. The system finalizes the dynamic process by performing another sliding step between the grey double-racks, introducing a 28 nm displacement at state VII. These seven states involving relative sliding and synchronous revolution are distinctly manifested by fluorescence data as shown in Fig. 4f. The corresponding theoretical calculations are presented in Supplementary Fig.21.

The self-assembly of multicomponent nanoarchitectures, which emulate the dynamic complexity of manmade macroscopic machinery outlines a fruitful avenue to sculpture matter on the nanoscale with tailored functionality. Alternatively, inspirations from biological machinery, for instance, motor proteins that can perform complicated mechanical motion in living cells still remain futuristic at this moment[30,31], as such natural wonders have developed through billions of years of evolution[32,33]. Interestingly, the only gearing system that has thus far been discovered in nature is from juvenile Issus, a plant-hopping insect, which has hind-leg joints, rotating like mechanical gears to synchronize its legs when jumping[34]. In contrary, human-designed gearing systems are versatile, ranging from simple, compound, reverted, to epicyclic gear trains[35]. In this regard, manmade machinery provides us a rich, yet more accessible resource than biological machinery, to build functional technomimetic analogues on the nanoscale. The DNA origami technique undoubtedly exemplifies a rigorous solution, because it affords an unprecedented level of accuracy to organize multiple components in a modular manner and meanwhile imparts dynamic complexity to the assembled nanoarchitectures. With such an engineering key at hand as well as the advent of deeper insights into how functional machinery can be built and mastered on the nanoscale, the objective of DNA-based molecular factories will not only be in our fantasies[36].

## Methods

**Design and preparation of the DNA origami filaments.**
The DNA scaffold strands (p7249 and p8064) were purchased from tilibit nanosystems. All other DNA strands were purchased from Sigma-Aldrich (high-performance liquid chromatography purification for the thiol-modified and dye-modified DNA. Reverse-phase cartridge purification for the staple, capture, blocking, and removal strands). The DNA origami structures were designed using caDNAno software[37]. To prevent non-specific aggregations, five thymine bases were added to the respective staple strands at the edge of the origami (design and sequence details can be found in Supplementary Figs 1, 12 and Supplementary Data 1, 2 and 3). The DNA origami structures were prepared by mixing 15 nM scaffold strands with 10 times of the staple strands and the capture strands in a buffer containing 0.5×TE (Tris, EDTA, pH = 8), 20 mM $MgCl_2$, and 5 mM NaCl. The mixture was then annealed as follows: 85°C for 5 min; from 65°C to 61°C, 1°C /5 min; from 60°C to 51°C, 1°C /90 min; from 51°C to 38°C, 1°C /20 min; from 37°C to 26°C, 1°C /10 min; held at 25°C. The annealed structures were purified with a filter device (100 kD, molecular weight cutoff, Amicon, Millipore) to remove excess staple and capture strands.

**Synthesis of the AuNRs.**
A two-step method was used to synthesize the AuNRs according to reference[38]. Synthesis of the seeds: 50 μL $HAuCl_4$ solution (2%, w/v) and 9.5 mL CTAB solution (100 mM) were added into a 20 mL flask. A 600 μL ice-cold $NaBH_4$ solution (10 mM) was then added to the flask under vigorous stirring for 2 min. The color of the solution quickly changed to yellowish brown. The solution was kept for 2 hours and then was used as nucleation seeds for the growth of AuNRs in the next step. Synthesis of the AuNRs: In a 20 mL flask, a 10 mL CTAB (hexadecyl-trimethyl-ammonium bromide, 100 mM) solution was first added, followed by addition of a 78 μL $HAuCl_4$ solution (2%, w/v). An 80 μL $AgNO_3$ solution (10 mM) was then added, changing the mixture to a light yellowish color. A 48 μL ascorbic acid solution (0.1 M) was injected with moderate shaking for 5 s. Upon addition of the ascorbic acid, the yellowish solution gradually became colorless. Finally, a 16 μL seed solution was injected and the flask was stirred for 1-2 min. The colorless solution turned to red, purple, and then brown. The resulting mixture was kept undisturbed at 30°C for 12 h for the AuNR growth. The product was isolated by centrifugation at 8,000 rpm for 10 min followed by removal of the supernatant. No size- or shape-selective fractionation was performed. The pellet was suspended in 100 μL water. The concentration was estimated by UV-Vis spectroscopy using extinction coefficient of $1\times10^9$ $M^{-1}cm^{-1}$ for the longitudinal plasmonic resonance of the AuNRs.

**Functionalization of the AuNRs with DNA.**
Functionalization of the AuNRs with thiolated DNA [5'-(ThiolC6)TTTTGACTTACC-3'] was carried out following the low-pH method[39]. A 4 mL AuNR solution (1 nM) was mixed with 40 μL sodium dodecyl sulfate (SDS, 1%), 400 μL 10× TBE, and 100 μL DNA (100 μM). Hydrochloric acid (HCl) was used to adjust the pH value of the mixture to 3. The disulfide bond in the thiolated oligonucleotides was reduced to monothiol using Tris(carboxyethyl) phosphine hydrochloride (TCEP) (20 mM, 1 hr) in water. The oligonucleotides were purified using size exclusion columns (G-25, GE Healthcare) to remove the small molecules. The purified DNA was added to the AuNR solution (OD~1) containing 0.01% (w/v) SDS with a molecular ratio of 4000:1. After incubation of the AuNRs with DNA for 1 h, a 5 M NaCl solution was added to bring the final concentration of NaCl to 500 mM. The solution was then gently shaken overnight.

Subsequently, the AuNR-DNA conjugates were centrifuged at 8,000 rpm for 25 min. The pellet was suspended in a 1 mL 0.5×TBE buffer containing 200 mM NaCl and the supernatant was discarded. The same centrifugation procedure was repeated three times to remove the excessive thiolated DNA completely. The final concentration of the AuNRs was estimated by UV-Vis spectroscopy using extinction coefficient of $1\times10^9$ $M^{-1}cm^{-1}$ for the longitudinal plasmonic resonance of the AuNRs.

**Synthesis of the AuNPs.**
AuNPs (10 nm) were synthesized using a two-step method[9]. A 1.25 mL $HAuCl_4$ solution (0.2%, w/v) was diluted in 25 mL double-distilled water and heated to boiling. A 1 mL sodium citrate solution (1%, w/v; containing 0.05% citric acid) was added to the flask under vigorous stirring. The solution in the flask was kept boiling for 5 min under stirring and then cooled down at room temperature.

**Surface modification of the AuNPs with BSPP.**
Bis(p-sulfonatophenyl)phenylphosphine dihydrate dipotassium salt (BSPP) (15 mg) was added to the AuNP solution (20 mL, OD ~ 1) and the mixture was shaken overnight at room temperature. Sodium chloride (solid) was added slowly while stirring until the solution color was changed from deep burgundy to light purple. The resulting mixture was centrifuged at 8,000 rcf for 30 min and the supernatant was removed with a pipette. The AuNPs were then resuspended in a 1 mL BSPP solution (2.5 mM). Upon mixing with 1 mL methanol, the mixture was centrifuged again at 8,000 rcf for 30 min. The supernatant was removed and the AuNPs were resuspended using ultra-pure water. The concentration of the AuNPs was estimated according to the optical absorption at 520 nm.

**Functionalization of the AuNPs with DNA.**
The AuNP-DNA conjugation was carried out according to Ding *et al*[40] with minor modifications. The disulfide bond in the thiol-modified oligonucleotides was reduced using tris(2-carboxyethyl)phosphine (TCEP) (100 mM, 1 h) in water. Thiol-modified oligonucleotides and BSPP modified AuNPs were then incubated at a molar ratio of DNA to AuNPs of 300:1 in a 0.5×TBE buffer solution for 20 h at room temperature. The concentration of NaCl was slowly increased to 300 mM in the subsequent 20 h in order to increase the thiolated DNA density on the AuNPs. The AuNP-DNA conjugates were then washed using a 0.5×TBE (tris-(hydroxymethyl)-aminomethan, borate, ethylenediaminetetraacetic acid) buffer solution in 100 kDa (MWCO) centrifuge filters to remove the free oligonucleotides. The concentration of the AuNP-DNA conjugates was estimated according to the optical absorption at 520 nm. Freshly prepared, fully coated AuNPs do not precipitate in a 0.5×TBE 10 mM $MgCl_2$ buffer.

**Self-assembly of the Au nanocrystals on DNA origami.**
For the independent revolution system, 10 times excess of the blocking strands 3, 6, 7, 8 and 9 were added to the purified DNA origami and incubated at room temperature for 0.5 h to block the footholds 3, 6, 7, 8 and 9 (attachment of the AuNRs at positions 1, 2 and 4, 5). For the synchronous revolution system, 10 times excess of the blocking strands 1, 4, 5 and 6 were added to the purified DNA origami and incubated at room temperature for 0.5 h to block the footholds 1, 4, 5 and 6 (attachment of AuNRs at positions 2 and 3). For the joint motion, 10 times excess of the blocking strands 3, 4, 5, 6, 7, 8, 9, 10, 11 and 12 were added to the purified DNA origami and subsequently incubated at room temperature for 0.5 h to block the corresponding footholds. The purified DNA-modified AuNRs or AuNPs were mixed with the DNA origami structures at a 10:1 molar ratio, and annealed by decreasing the temperature from 38 °C to 25 °C at a rate of 1 °C per 60 min. An agarose gel purification step (1% agarose gel in a 0.5×TBE buffer with 11 mM $MgCl_2$) was used to purify the successfully assembled product.

**TEM characterization.**
The DNA origami structures were imaged using Philips CM 200 TEM operating at 200 kV. For imaging, the purified samples were deposited on freshly glow-discharged carbon/formvar TEM grids. Before depositing the sample solution, the grids were treated by negative glow discharge. After 10 min deposition, TEM grids were treated with a uranyl formate solution (2%) for negative staining of the DNA structures. Uranyl formate for negative TEM staining was purchased from Polysciences, Inc.. Averaged TEM images were obtained using EMAN2 software[41].

**Fluorescence spectroscopy**
Fluorescence spectra were measured using a Jasco-FP8500 Fluorescence Spectrometer with a quartz SUPRASIL ultra-micro cuvette (path length, 10 mm). All measurements were carried out at room temperature in a buffer after agarose gel purification (0.5×TBE buffer with 11 mM $MgCl_2$, pH = 8). For the *in situ* fluorescence measurements, a 120 µL solution containing ~1 nM of the structures at the initial configuration was used. The fluorescence emissions at 578 nm and 663 nm were monitored using the dual-wavelength time-scan acquisition mode and a data pitch of 10 s. The excitation wavelength were 550 nm and 647 nm for the ATTO 550 and ATTO 647N, respectively. Respective blocking and removal strands were added to enable programmable rotations.

**Numerical simulations**
All simulations were carried out using the software COMSOL Multiphysics, which is a commercially available finite-element solver. The AuNR was modeled as a cylinder with a diameter of 10 nm and a length of 35 nm. In order to approximate the shapes visible in the TEM images, the ends of the cylinder were not taken as flat surfaces, but instead constructed as half spheres with a radius of 5 nm. The permittivity of gold was taken from the published paper[42]. The refractive index of the surrounding medium (water) was 1.332. The intrinsic quantum yields (ATTO 550: $q_0 = 0.8$, ATTO 647N: $q_0 = 0.65$) and the intrinsic emission spectra of the two fluorophores were taken as specified by the supplier (www.atto-tec.com).

**Data availability.** The data that support the plots within this paper and other findings of this study are available from the corresponding author upon reasonable request.

## Acknowledgements

This project was supported by the Sofja Kovalevskaja grant from the Alexander von Humboldt-Foundation, the Volkswagen foundation, and the European Research Council (ERC Dynamic Nano) grant. M.U. acknowledges the financial support by the Carl-Zeiss-Stiftung. We thank Marion Kelsch for assistance with TEM. TEM images were collected at the Stuttgart Center for Electron Microscopy.

## Author contributions

N.L. and P.F.Z. conceived the project. P.F.Z and M.U. designed the DNA origami nanostructures. P.F.Z. performed all the experiments. S.B. and T.W. carried out the theoretical calculations. P.F.Z., S.B., and N.L. wrote the manuscript. A.K. made helpful suggestions. All authors discussed the results and commented on the manuscript.

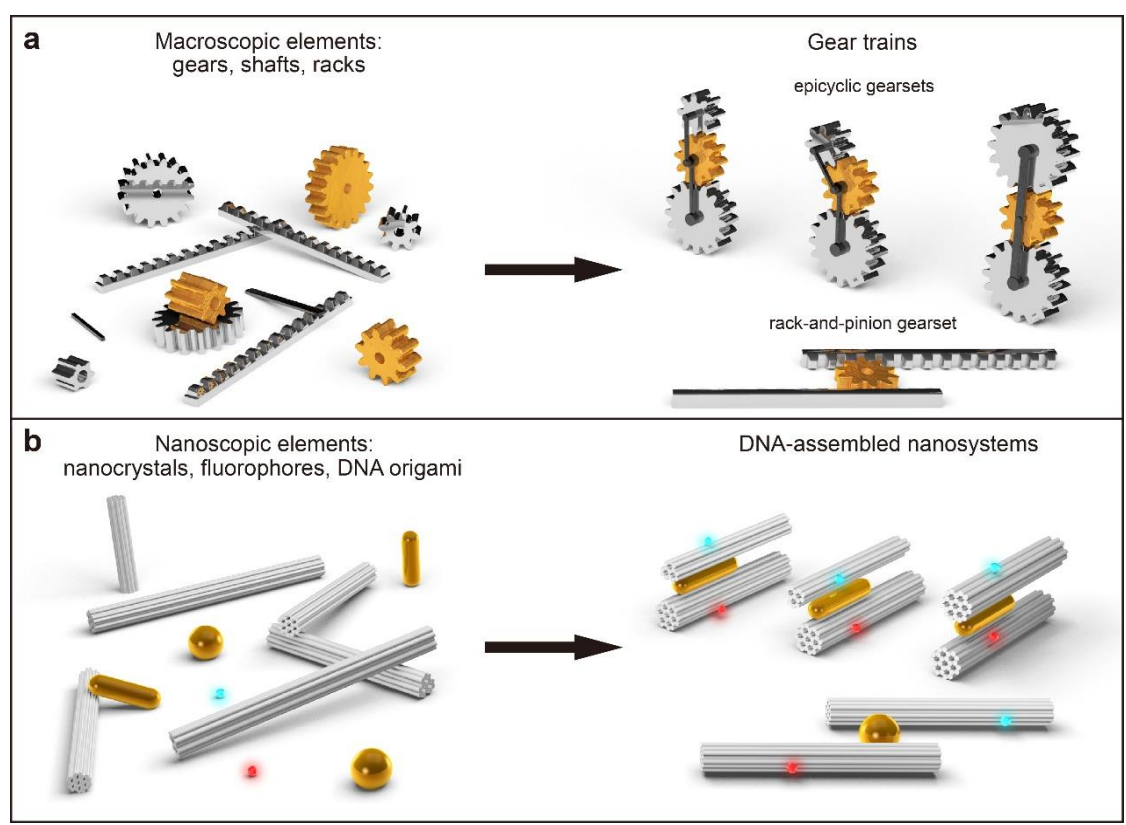

**Figure 1 | Schematic of the macroscopic and nanoscopic systems with regulated and coordinated motion. a**, Macroscopic elements such as gears, shafts, and racks are assembled together to form gear trains. **b**, Nanoscopic elements including DNA origami filaments, fluorophores, and metallic nanocrystals are assembled together to form the technomimetic analogues of the gear trains for executing independent, synchronous or joint motion.

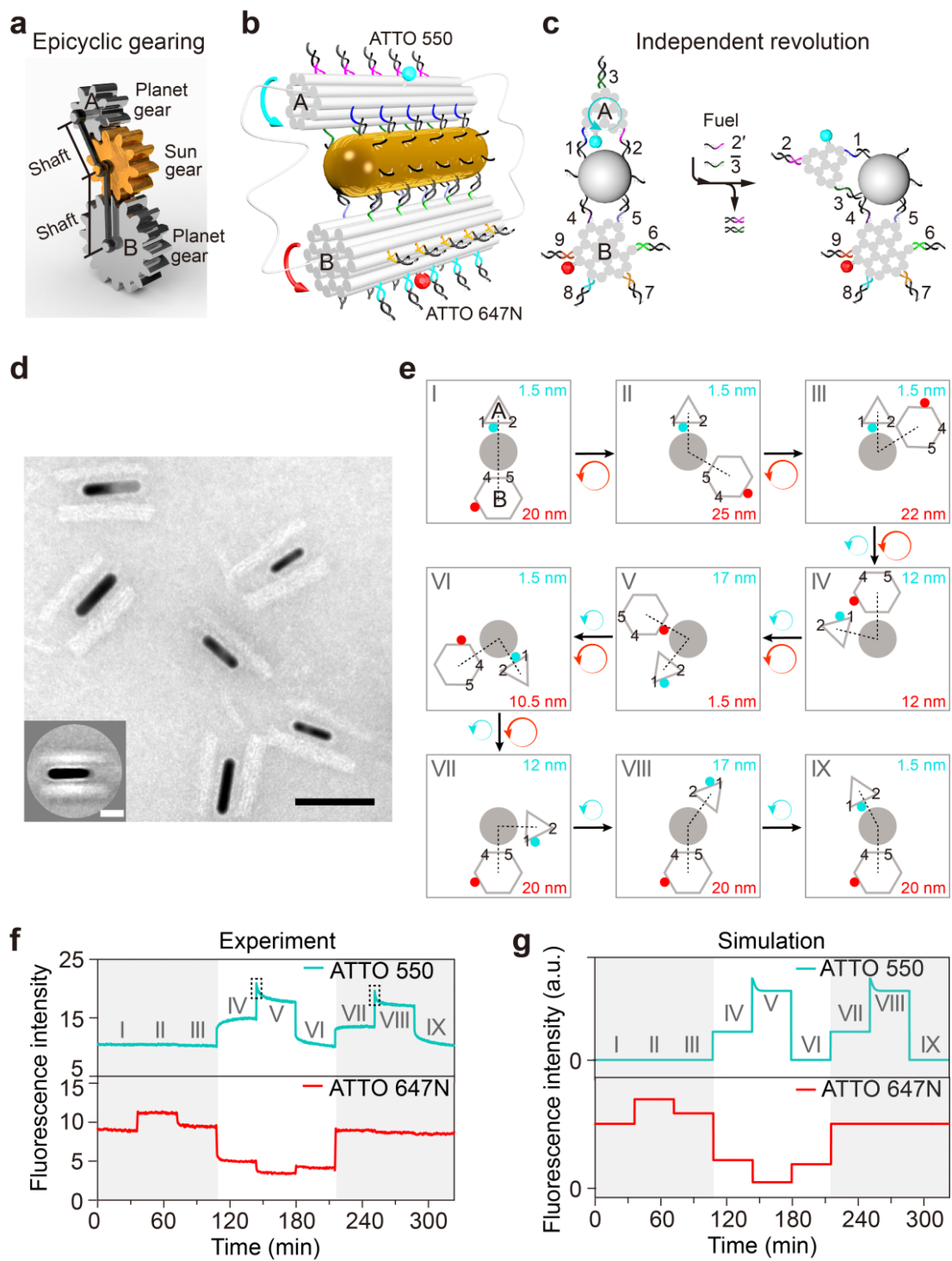

**Figure 2 | Independent revolution. a**, Epicyclic gearset, in which two planet gears (A and B, grey) of different diameters mounted on a sun gear (brown) using two shafts can independently revolve around the sun gear. **b**, DNA-assembled hybrid nanosystem, in which two origami filaments (A, 13-helix and B, 23-helix) anchored on the side-surface of a gold nanorod (AuNR) through DNA hybridization can independently revolve around the AuNR. Two fluorophores (ATTO 550 and ATTO 647N) are tethered on filaments A and B, respectively, allowing for *in situ* optically monitoring the revolution process. **c**, Cross-section view of the system. Three rows of footholds (coded 1–3) evenly separated by 120° and six rows of footholds (coded 4–9) evenly separated by 60° are extended from filaments A and B, respectively.

Upon addition of blocking and removal strands, toehold-mediated strand displacement reactions enable counterclockwise revolution of filament A around the AuNR by rotation, while filament B keeps its anchoring position on the AuNR. **d**, TEM image of the AuNR-origami structures before rotation. Scale bar, 50 nm. Inset: averaged TEM image. Scale bar, 20 nm. **e**, Representative route for an independent revolution process, comprising nine distinct states (I–IX). The positions of the two fluorophores and their relative distances to the AuNR surface along the respective radial directions are given for each state. Experimental measurements (**f**) and theoretical calculations (**g**) of the fluorescence intensities of ATTO 550 (blue) and ATTO 647N (red) during the independent revolution process from I to IX. Kinks in fluorescence are observed, when filament A revolves around the AuNR, transiting from IV to V and VII to VIII, respectively, as highlighted by the black-dashed frames.

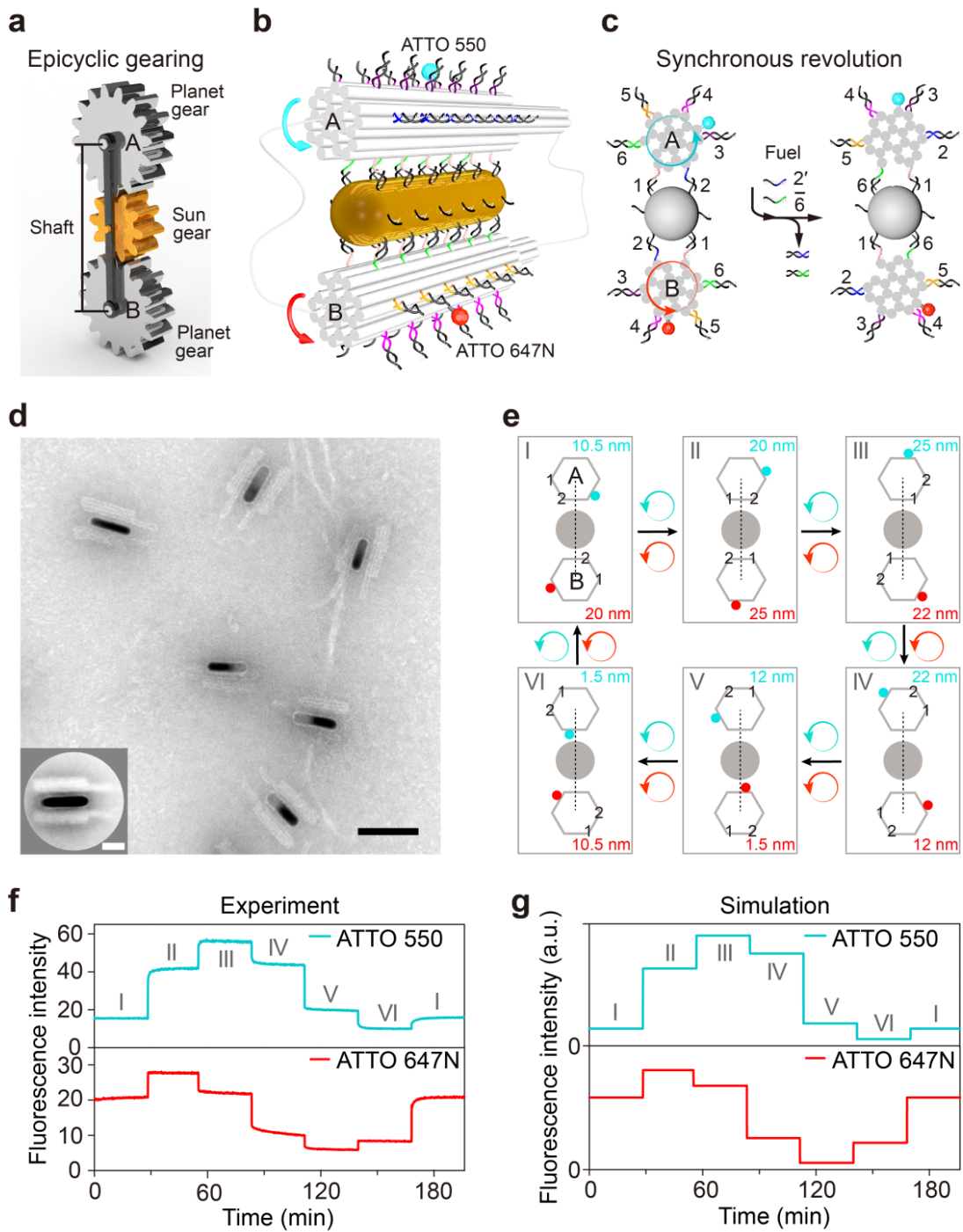

**Figure 3 | Synchronous revolution. a**, Epicyclic gearset, in which two planet gears (A and B, grey) of the same diameter mounted on a sun gear (brown) using one single shaft can synchronously revolve around the sun gear. **b**, DNA-assembled hybrid nanosystem, in which two origami filaments (A and B, 23-helix) anchored on the side-surface of a AuNR through DNA hybridization can synchronously revolve around the AuNR. Two fluorophores (ATTO 550 and ATTO 647N) are tethered on filaments A and B, respectively. **c**, Cross-section view of the system. Six rows of footholds (coded 1–6) evenly separated by 60° are extended from each filament. Upon addition of the same set of the DNA fuels, toehold-mediated strand displacement reactions enable synchronous revolution of the two filaments around the AuNR. **d**, TEM image of the AuNR-origami structures before rotation. Scale bar, 50 nm.

Inset: averaged TEM image. Scale bar, 20 nm. **e**, Representative route for a synchronous revolution process, comprising six distinct states (I–VI). The positions of the two fluorophores and their relative distances to the AuNR surface along the respective radial directions are given for each state. Experimental measurements (**f**) and theoretical calculations (**g**) of the fluorescence intensities of ATTO 550 and ATTO 647N, respectively, during the synchronous revolution process from I to VI.

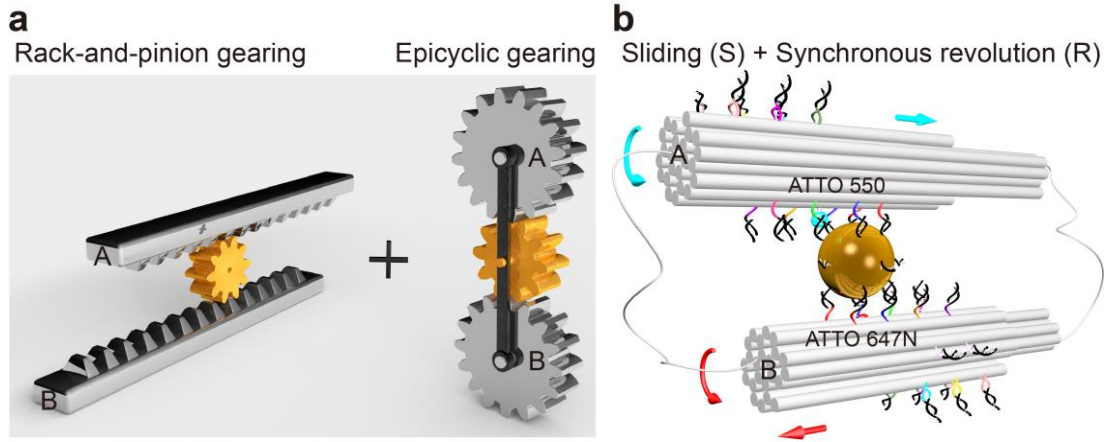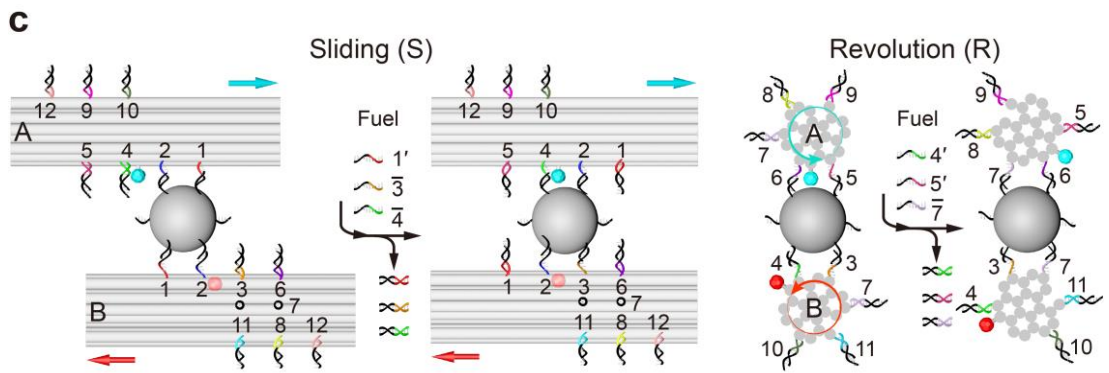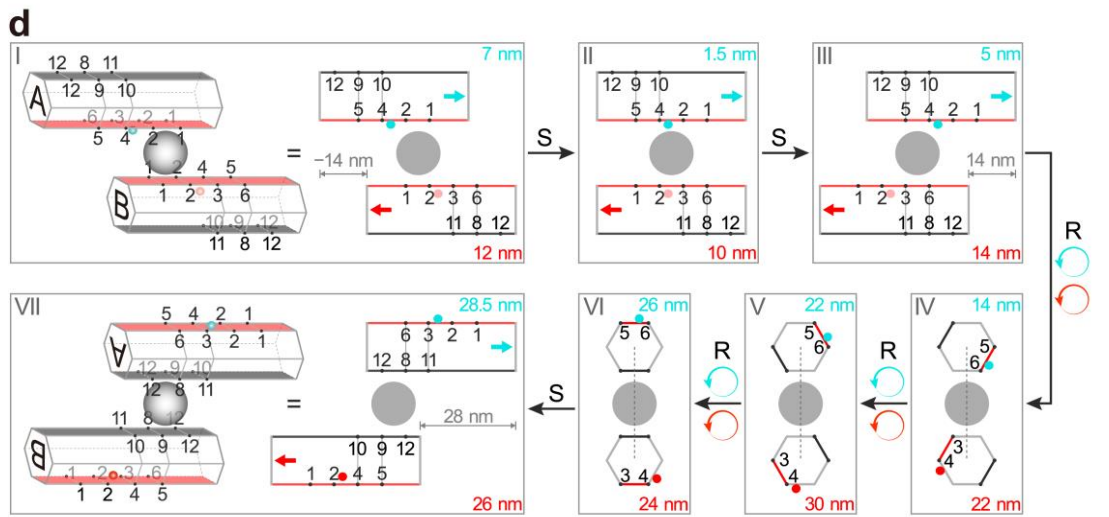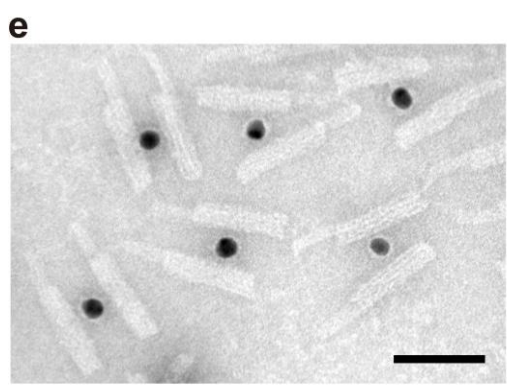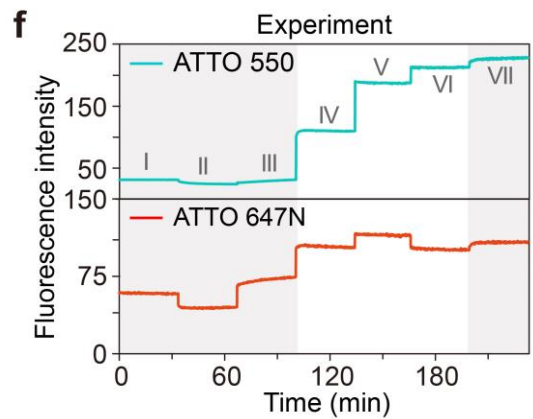

**Figure 4 | Joint motion. a**, Combination of relative sliding from rack-and-pinion gearing and synchronous revolution from epicyclic gearing. **b**, DNA-assembled hybrid nanosystem, in which two origami filaments (A and B, 23-helix) anchored on a gold spherical nanoparticle (AuNP, 10 nm) through DNA hybridization. Twelve rows of footholds are extended from the filaments. **c**, Relative sliding and synchronous revolution enabled by toehold-mediated strand displacement reactions upon addition of corresponding DNA fuels. **d**, Representative route for joint motion, comprising seven distinct states (I–VII). The positions of the two fluorophores and their relative distances to the AuNP surface along the respective radial directions are given for each state. The two different double-racks for relative sliding are indicated by parallel planes in red and grey, respectively. **e**, TEM image of the AuNP-origami structures at state I. Scale bar, 50 nm. **f**, Experimental measurements of the fluorescence intensities of ATTO 550 and ATTO 647N, respectively, during the joint-motion process from I to VII.